\documentclass[aps]{revtex4}

\usepackage{amssymb}
\usepackage[tbtags]{amsmath}

\newcommand{\SSC}{\scriptscriptstyle}
\newcommand{\abs}[1]{\lvert#1\rvert}

\begin{document}
\title{Lorentz Force Correction and Radiation Frequency Property\\
 of Charged Particles in Magnetic Dipole}
\author{Ji Luo\footnote{Author thanks Fang-an Wang for valuable discussion.},
Chuang Zhang, Bo Liu}
\email{luoji@mail.ihep.ac.cn}
\affiliation{Accelerator Center, Institute of High Energy Physics, Beijing, China}

\date{\today}

\begin{abstract}
By concern of compression of charge density field, the corrected Lorentz force formula
and consequent inference is presented. And further radiation frequency property of an
individual charge density field in magnetic dipole is analyzed respectively for radiant
property of the charged particle and the emitted electromagnetic wave transfer property
between the moving radiant source and observer. As results, the behavior and radiation
frequency property of the electron beam in magnetic dipole is interpreted upon the
individual's behavior and property. At final, the potential application is put forward
for wider interest.
\end{abstract}

\maketitle

\section{Introduction}
In view point of electrical charge density field\ \cite{luoji:chargedensityfield}, Lorentz interaction of
individual charge and magnetic field can happen only in a specified space ---
overlapped volume of the charge density field and magnetic field. When the volume is in
compressed status or varies with a change of the charge internal potential energy due
to energy exchange, the Lorentz force will be affected since the magnetic action amount
is proportion to the volume; consequently the centrifugal force's variation will result
in the angular velocity's variation and radiant power's of the charged particle. By
considering the charge's internal structure and compressed status, Lorentz force
correcting factor is studied. In addition, for the Lorentz force to work for causing
electromagnetic radiation, there will be a corresponding property on the radiation
frequency of a single particle during a single pass; however on the radiation property
involving the particle's radiant frequency, observer detecting frequency and radiant
power, currently relevant study is not complete and explicit yet in some points for neglecting
the charge's micro-structure and undistinguishing the particle charge itself's
radiation in magnetic field from detecting radiation at observer
location\ \cite{StanleyHumphries:pcpa,Alexander:ape,Edwards:iphea}, while it is one of our studying subjects following.
Further based on the new results of Lorentz force correction and radiation frequency
property for single charged particle in magnetic dipole, we give a qualitative
interpretation on the relevant behavior or property of collectively charged particles
or macro beam element during it passes through a magnetic dipole, and discuss verifying
method on our theorem as well as important application.

\section{Derivation of the Correct Relation and Inference}
\subsection{Corrected Formula of Lorentz Force}
For a micro charge element within a charge density field, where
$dq_i=\rho_i(t)dV_i(t)=\rho_{_{0}i}(r_{_{0}ic})dV_{0i}(r_{_{0}ic})$ and $\iiint\limits_{V(
t)}\rho[r_{ic}(t)]dV=\iiint\limits_{V_0}\rho_{_{0}}(r_{_{0}ic})dV_0(r_{_{0}ic})=q$ (Ref.\ %
\cite{luoji:chargedensityfield}), there exists a corrected Lorentz force exerted on the $dq_i$ as
below:
\begin{equation}
\label{eq201} d\vec f\;[r_{ic}(t)]=dq_i\vec v\;[r_{ic}(t)]\times \vec
B[r_{ic}(t)]k_i[r_{ic}(t)]
\end{equation}
\[
\hfill k_i(t)=\frac{dV_i}{dV_0}=e^{-\eta_i(t)}\qquad\eta_i\in[0,+\infty)\qquad
k_i\in(0,1]
\]

Integrating Eq.\ (\ref{eq201})
\begin{equation}
\label{eq202}\begin{split} \vec f\;[r_{cc}(t)]&=\iiint\limits_{V(\dot
t)}\rho[r_{ic}(t)]\vec v\;[r_{ic}(t)]\times\vec B[r_{ic}(t)]k[r_{ic}(t)]dV[r_{ic}(t)]\\
&=\vec v_{cc}(t)\times\vec B(t)\bar k(t)q
\end{split}
\end{equation}
where $\bar k(t)$ is average compression ratio and $\bar k(t)=e^{-\bar \eta(t)}$~\cite{luoji:chargedensityfield}.

\subsection{Consequent Inferences}
\label{sec202}
\renewcommand{\thesubsubsection}{\Alph{subsubsection}}

\subsubsection{Angular velocity relation of a charged particle with mass $m_q$ in
uniform magnetic dipole}

From
\begin{widetext}
\begin{equation}
\label{E:1}
\rho(t)\frac{d\Vec n_{\rho}(t)}{dt}=\rho(t)\omega(t)\Vec n_{\bot\rho}(t)
=\rho(t)\omega_{\rho}(t)\Vec n_{v}(t)=v(t)\Vec n_v(t)
\end{equation}
Take derivative Eq.~\eqref{E:1} with respect to $t$
\begin{equation}
\label{E:2}
\left[\dot \rho(t)\omega_{\rho}(t)+\rho(t)\dot \omega_{\rho}(t)\right]\Vec n_v-\rho\omega^2_{\rho}(t)\Vec n_{\rho}%
=\dot v(t)\Vec n_v+v(t)\omega_{\rho}(t)\Vec n_{\bot v}
\end{equation}
\end{widetext}

From Eq.~\eqref{eq202} and Eq.~\eqref{E:2} there
\begin{equation}
\label{eq203}  m_q\rho(t)\omega_{\rho}^2(t)=\bar k(t)qv(t)B(t)
\end{equation}
where $\rho(t)$ --- curvature radius of the charge center of a charged particle in magnetic dipole,
$\Vec n_{\rho}(t)$, $\Vec n_v(t)$ is unit vector of $\Vec \rho(t)$, $\dot \Vec r(t)$ respectively,
$\Vec\rho=\rho\Vec n_{\rho}$, $\dot \Vec r=v \Vec n_v$; $m_q$ --- rest mass of charged particle.

There
\begin{equation}
\label{eq203b}\tag{\ref{eq203}$'$} \omega_{\rho}(t)=\frac{\bar
k(t)qB}{m_q}=\frac{qB}{m_q}\frac{1}{e^{\bar\eta(t)}}
\end{equation}

That is, the particle's angular or revolution frequency is inversely proportional to
the internal potential energy of the particle or to the compression status of the
charge density field.

And
\begin{equation}
\label{eq203c}\tag{\ref{eq203}$''$}\frac{d\omega_{\rho}(t)}{dt}=\frac{qB}{m_q}\frac{d\bar
k(t)}{dt}=-\frac{qB}{m_q}e^{-\bar\eta(t)}\cdot\frac{d\bar\eta(t)}{dt}=-\omega_{\rho}(t)\frac{d\bar\eta(t)}{dt}
\end{equation}
\subsubsection{Resultant force exerted on the charged particle and derivation of the rate
of curvature radius of charge center}
\begin{equation} \label{eq204}
\begin{split}
\vec f_{q}(t)&=m_q\dot v(t)\Vec n_v+m_q v(t)\omega_{\rho}(t)\Vec n_{\bot v}\\
&=m_q(\dot \rho\omega_{\rho}+\rho\dot\omega_{\rho})\Vec n_v+m_q \rho \omega_{\rho}^2\Vec n_{ca}\\
&=m_q(\dot \rho\omega_{\rho}+\rho\dot\omega_{\rho})\Vec n_v+\bar k(t)q{\dot{\Vec r}}(t)\times\vec B\\
&=f_{v}(t)\vec n_v(t)-f_{\rho}\vec n_{\rho}(t)
\end{split}
\end{equation}

In frame fixated on the particle, according to the fact that the rate of line velocity,
denoted as $a_{\SSC
//}(t)$, results from centrifugal acceleration $a_{\bot}(t)$, in other words $a_{\SSC //}(t)$ is
determined by $a_{\bot}(t)$ and its value is just profile of the centrifugal
acceleration on $\vec n_{\bot \rho}$ direction.

So there
\begin{equation}
\label{eq205a} \frac{a_{\SSC
//}(t)}{a_{\bot}(t)}=\frac{\dot v(t)}{a_{\bot}(t)}=\pm^\ast\!\!\tau \omega_{\rho}(t)
\end{equation}
where $^\ast\tau=1\text{ second}$, and as $\dot v>0$, take the symbol as ''$+$''; as $\dot v<0$, take the symbol as ''$-$''.

to substitute $a_{\bot}(t)=v(t)\omega_{\rho}(t)$ to
Eq.\ (\ref{eq205a}), then
\begin{equation}
\label{eq205b}\tag{\ref{eq205a}$'$}
\dot v=\pm^\ast\tau v(t)\omega^2_\rho(t)=\begin{cases}
^\ast\tau v\omega^2_\rho    \qquad \dot v>0\\
-^\ast\tau v \omega^2_\rho  \qquad \dot v<0
\end{cases}
\end{equation}

Resolve Eq.\ (\ref{E:2}) by using Eq.\ (\ref{eq205b})\\
thus \begin{equation}\label{eq205c}
 \dot
\rho(t)=\frac{\dot v}{\omega_{\rho}}-\rho\frac{\dot {\bar k}}{\bar k}=\pm^\ast\!\tau\frac{\bar k(t)qBv(t)}{m_q}-\rho\frac{\dot {\bar k}}{\bar k}
\end{equation}

\subsubsection{Radiation power formula}
\renewcommand{\labelenumi}{(\roman{enumi})}
\begin{enumerate}
\item Radiant power formula of an electron in magnetic dipole

In case $\dot v<0$, $\dot{\bar k}>0$, by using Eq.\ (\ref{eq204}) or $P_{k\ell}=\Vec f_{q}(t)\cdot \dot{\Vec r}(t)= m_qv\dot v$ and
$\dot v=-^\ast\tau v\omega^2_\rho$, $E_p=\mu e (\ln\bar k^{-1}+\bar k-1)$ [Ref.~\cite{luoji:chargedensityfield}]\\
\begin{equation}
\label{eq206}
\begin{split}
P_{t\ell}(t)&=-P_r(t)\\
&=P_{k\ell}(t)+P_{p\ell}(t)\\
&=m_qv\dot v+\frac{dE_p(t)}{dt}\\
&=\frac{d\left[\frac{m_q}{2}v^2(t)+E_p(t)+E_0\right]}{dt}\\
&=-^\ast\tau m_q v^2 \omega^2_\rho-\mu e \dot{\bar k}\frac{1-\dot{\bar k}}{\bar k}\\
&=-^\ast\!\tau\frac{e^2 B^2 \bar k^2(t) v^2(t)}{m_q}-\mu e \dot {\bar k}\frac{1-\bar k}{\bar k}
\end{split}
\end{equation}
\begin{equation}
\label{E:radiantpower}
P_r=^\ast\tau m_qv^2\omega^2_\rho+\mu e\dot{\bar k}\frac{1-\bar k}{\bar k}
\end{equation}

where $P_r$ --- radiant power; $P_{t\ell}$ --- power of electron's total energy loss; $P_{k\ell}$ ---
power of electron's kinetic energy loss; $P_{p\ell}$ --- power of electron's potential energy loss.
$E_p$ --- internal potential energy of electron's charge density field;
$E_0$ --- electron's intrinsic energy, and $E_0 = constant$ Ref.~\cite{luoji:chargedensityfield}.

\item Detecting power

Ref.\ \cite{luoji:transferrelation} and suppose $\mathcal J_{exc}(t')=0$, $\mathcal J_l(t')=0$; there
\begin{equation}
\label{eq206a} P_0(t')=\mathcal J(t')=\mathcal
J(t)\frac{dt}{dt'}=P_r(t)\left[1-\frac{dT(t')}{dt'}\right]
\end{equation}

\end{enumerate}
\section{Radiation Frequency Property for Single Pass of Single Particle}
\subsection{Charge's radiant frequency property}
\label{sec301}

From
\[
\dot \rho(t)=\bar k(t)\lambda_0\nu(t)=\pm^\ast\!\!\tau\frac{\bar k(t)qBv}{m_q}+\rho\frac{\dot{\bar k}}{\bar k}
\]
get
\begin{equation}
\label{eq207}\nu(t)=\pm\frac{^\ast\!\tau qBv(t)}{m_q\lambda_0}+\rho\frac{\dot {\bar k}}{\bar k^2\lambda_0 }
\end{equation}
and
\begin{equation}
\label{eq208}\frac{d\nu(t)}{dt}=^\ast\!\!\tau\frac{qB}{m_q\lambda_0}\dot v(t)+\frac{d(\rho\frac{\dot {\bar k}}{\bar k^2\lambda_0})}{dt}
\end{equation}
and
\begin{equation}
\begin{split}
\Delta \nu_r&=\left[\pm\frac{^\ast\!\tau qBv(t)}{m_q\lambda_0}+\frac{1}{\lambda_0}%
\rho(t)\frac{\dot {\bar k}(t)}{\bar k^2(t)}\right]_{t_a}^{t_b}\\
&=\pm\frac{^\ast\!\tau qB}{m_q\lambda_0}\abs{v(t_a)-v(t_b)}+\frac{1}{\lambda_0}\left[\rho(t_a)%
\frac{\dot{\bar k}(t_a)}{\bar k^2(t_a)}-\rho(t_b)\frac{\dot{\bar k}(t_b)}{\bar k^2(t_b)}\right]
\end{split}
\end{equation}
where $\lambda_0$ --- diameter of intrinsic charge density field

Here there is continuous electromagnetic compression status in horizontal direction,
$\vec n_\rho(t)$; it is resulted from a resultant electrical field action consisted of two
induced electrical field in opposite directions; one in centrifugal direction for
Lorentz deflection or Lorentz force to work, the other in opposite direction of the
displacement for created radiant magnetic field to oppose against the Lorentz work and
convert the work into horizontal polarized electromagnetic radiation in tangential
direction of the particle's trajectory.

\subsection{Observer detecting frequency property}
Ref.\ \cite{luoji:timefunction} there is a frequency transfer relation between the radiant source
and observer as following
\begin{equation}
\label{eq209}\nu(t')=\nu(t)\frac{dt}{dt'}=\nu(t)\left[1-\frac{dT(t')}{dt'}\right]
\end{equation}
here $\frac{dT(t')}{dt'}<0$, so $1-\frac{dT(t')}{dt'}>1$ and $\nu(t')>\nu(t)$

Specifically in optical non-dispersion medium,
\begin{equation}
\label{eq210} \nu(t')=\nu(t)\left[1+\frac{v(t')}{c}\right]
\end{equation}
here suppose $v_o(t')=0$, where $v_o(t')$ --- speed of observer.

In optical dispersion medium
\begin{equation}
\label{eq211}\nu(t')=\nu(t)\left[1+\frac{v(t')+a_{sis}(t)T(t)}{c-a_{sis}(t)T(t)}\right]
\end{equation}

Furthermore Ref.\ \cite{luoji:redshift}
\begin{widetext}
\begin{equation}
\label{eq212}
\begin{split}
a_{sis}(t)=a_{im}(t)
&=\frac{dv_{im}(t)}{dt}=\frac{dv_{im}}{d\omega}\frac{d\omega_{im}(t)}{dt}+\frac{dv_{im}}{dn}\cdot\frac{dn}{dr_s}v(t)\\
&=\frac{dv_{im}}{d\omega}\left[\frac{d\omega_s(t)}{dt}+\rho_{s\!p}(t,n)a_{sm}(t)\right]+\frac{dv_{im}}{dn}\cdot\frac{dn}{dr_s}v(t)\\
&=\frac{dv_{im}}{d\omega}\left[2\pi\frac{d\nu(t)}{dt}+\rho_{s\!p}(t,n)\dot v(t)\right]+\frac{dv_{im}}{dn}\cdot\frac{dn}{dr_s}v(t)
\end{split}
\end{equation}
\begin{tabular}{rrcl}
\indent where & $n$ &---&refractive index of the dispersion medium\\
&$v_{im}$&---&velocity of electromagnetic wave with respect to medium\\
&$\rho_{s\!p}(t,n)$&---&phase density of radiant source at instant $t$ and in medium\\
&&&with refractive index $n$
\end{tabular}
\end{widetext}

As the medium has same refractive index, then from Eq.\ (\ref{eq211})
\begin{equation}
\label{eq213}a_{sis}(t)=\frac{dv_{im}}{d\omega}\left[2\pi\frac{d\nu(t)}{dt}+\rho_{s\!p}(t,n)\dot v(t)\right]
\end{equation}

From Eq.\ (\ref{eq211}) and (\ref{eq213}), it is known that in even dispersion medium
detected frequency at observer location is related to various information about the
charged particle and signal transfer time $T(t)=T(t')$.

\subsection{Observing frequency width}
According to Eq.\ (\ref{eq209}) as for radiant source's highest frequency $\nu_h(t_h)$
and lowest frequency $\nu_l(t_l)$ there exist the corresponding relations
\[
\nu_h(t_h')=\nu_h(t_h)\left[1-\left.\frac{dT(t')}{dt'}\right|_{t_h'}\right]
\]
and
\[
\nu_l(t_l')=\nu_l(t_l)\left[1-\left.\frac{dT(t')}{dt'}\right|_{t_l'}\right]
\]
then
\begin{widetext}
\begin{equation}
\label{eq214}\nu_h'(t_h')-\nu_l'(t_l')=\nu_h(t_h)\left[1-\left.\frac{dT(t')}{dt'}\right|_{t_h'}\right]%
-\nu_l(t_l)\left[1-\left.\frac{dT(t')}{dt'}\right|_{t_l'}\right]
\end{equation}

Suppose $\left.\frac{dT(t')}{dt'}\right|_{t_h'}=\left.\frac{dT(t')}{dt'}\right|_{t_l'}$
and using Eq.\ (\ref{eq207})
\begin{equation}
\label{eq215}
\nu_h'-\nu_l'=(\nu_h-\nu_l)\left[1-\left.\frac{dT(t')}{dt'}\right|_{t_l'}\right]=%
\Delta \nu_r\left[1-\left.\frac{dT(t')}{dt'}\right|_{t_l'}\right]
\end{equation}

Therefore for non-dispersion medium
\begin{equation}
\label{eq216}
\nu_h'-\nu_l'=(\nu_h-\nu_l)\left[1+\frac{v(t')}{c}\right]=2\Delta \nu_r=2\left\{
\pm\frac{^\ast\!\tau qB}{m_q\lambda_0}\abs{v(t_a)-v(t_b)}+\frac{1}{\lambda_0}\left[\rho(t_a)%
\frac{\dot{\bar k}(t_a)}{\bar k^2(t_a)}-\rho(t_b)\frac{\dot{\bar k}(t_b)}{\bar k^2(t_b)}\right]\right\}
\end{equation}
for even or uniform dispersion medium, and combine Eq.\ (\ref{eq208}), (\ref{eq213}) with
Eq.\ (\ref{eq215})
\begin{equation}
\label{eq217}
\begin{split}
\nu_h'-\nu_l'
&=(\nu_h-\nu_l)\left[1+\frac{v(t')+a_{sis}(t)T(t)}{c-a_{sis}(t)T(t)}\right]\\%
&=\Delta\nu_r\left\{1+\frac{c+\frac{dv_{im}}{d\omega}%
\left[^\ast\!\tau\frac{2\pi qB}{m_q\lambda_0}\dot v+\frac{d(\rho\frac{\dot {\bar k}}{\bar k^2\lambda_0})}{dt}+\rho_{s\!p}(t,n)\dot v\right]T(t)}{c-\frac%
{dv_{im}}{d\omega}\left[^\ast\!\tau\frac{2\pi%
qB}{m_q\lambda_0}\dot v+\frac{d(\rho\frac{\dot {\bar k}}{\bar k^2\lambda_0})}{dt}+\rho_{s\!p}(t,n)\dot v\right]T(t)}\right\}
\end{split}
\end{equation}
\end{widetext}

\section{Discussion}
\subsection{Lorentz force, in a complete inertial frame, is particle's total energy
related}

For a single charged particle, its energy $E_q(t)=\frac{m_q}{2}v^2(t)+\bar \mu
q[\ln \bar k^{-1}(t)+\bar k(t)-1]+E_{q0}=\frac{m_q}{2}v^2(t)+\bar \mu q[\bar
\eta(t)+e^{-\bar \eta (t)}-1]+E_{q0}$ (Ref.\ \cite{luoji:energydifferential}) relates to corrected Lorentz
force as centrifugal force $\vec f_{\rho}(t)=\bar k(t)q\vec v(t)\times\vec B(t)$; where
$\vec v(t)$ reflects that the force is particle's kinetic energy related, and
$\bar k(t)=e^{-\bar\eta(t)}$ shows the corrected force is also particle charge's
internal potential energy related.

For energy loss process of a charged particle in magnetic field, there
$\frac{\dot v}{a_{\bot}}=-^\ast\tau \omega_{\rho}(t)$, or $\dot v=-^\ast\tau v\left(\frac{\bar k
q B}{m_q}\right)^2$. While for energy increase process of a charged particle in magnetic field
from external electrical field, where its intensity is $\mathbf E(t)$, $\frac{\dot a_{\bot}}{\dot v}=\omega_{\rho}(t)$,
or $\dot a_{\bot}=(v\omega_{\rho})'=\dot v \omega_{\rho}=\frac{\mathbf E(t)q}{m_q}\frac{\bar k q B}{m_q}$.

\subsection{Interpretation on behavior and radiation property of electron beam in
magnetic dipole}
\subsubsection{Beam's horizontal size decreasing}
This is because individual's charge density field is compressed by two opposite
electrical fields (Ref.\ \ref{sec301}) along centrifugal direction or $\vec n_r(t)$.
The compression contributes to decreasing space charge effect along this direction.
\subsubsection{Decreasing of beam's longitudinal kinetic energy spread}
This is because along $\vec n_{\bot r}(t)$ there exists $\frac{\partial
v(t,z)}{\partial z}<0$; thus along the longitudinal direction beam's internal
force increase, that help to improve the kinetic energy spread and increase internal
potential energy.
\subsubsection{Observer's frequency width is about two times of source itself's
frequency width}

This is because there is derivative of frequency transfer relation Eq.\ (\ref{eq209})
\begin{equation}
\label{eq218}
\begin{split}
\frac{d\nu(t')}{dt'}&=\frac{d\nu(t)}{dt}\left(\frac{dt}{dt'}\right)^2+\frac{d\nu(t)}{dt}\frac{d^2t}{dt'^2}\\
&=\frac{d\nu(t)}{dt}\left\{\left[1-\frac{dT(t')}{dt'}\right]^2-\frac{d^2T(t')}{dt'^2}\right\}
\end{split}
\end{equation}
or
\begin{equation}
\label{eq218b}\tag{\ref{eq218}$'$}
d\nu(t')=d\nu(t)\left\{1-\frac{dT(t')}{dt'}-\left[%
1+\frac{dT(t)}{dt}\right]\frac{d^2T(t')}{dt'^2}\right\}
\end{equation}

For example in non-dispersion medium, $\frac{d^2T(t')}{dt'^2}=0$,
$\frac{dT(t')}{dt'}=-\frac{v(t')}{c}\doteq -1$

So from Eq.\ (\ref{eq218b}), there
\begin{equation}
\label{eq219} d\nu(t')\backsimeq 2d\nu(t)
\end{equation}

For beam the observer frequency width will be wider than single's due to energy spread
of the electron beam.

\subsection{Application}
A. Corrected Lorentz force formula can be used to analyze the deflecting angle of the charge
particles with different energy in magnetic dipole. Consequent gyral frequency formula can
be used to analyze circling frequency of charge
particle in storage ring or cyclotron directly, without supposing mass's
increasing or decreasing of particle as particle's energy varies.
(Ref.\ Eq.\ (\ref{eq203b}) (\ref{eq203c}))

B. The radiation frequency and power property can be applied to analyze
radiant frequency and power property of synchrotron light sources.
The property can also be used in analysis of light source of high red-shift astronomic
observation.

\section{Conclusion}
Lorentz force in magnetic field is corrected by charge field's compression factor which
manifests charge density field's internal potential energy status. The radiation
frequency property of individual charged particle in magnetic field is consisted of two
characteristics, radiant frequency property due to Lorentz deflection radiation and
detecting or observing frequency property which is determined by time function of
signal transportation between time domains of radiant source and observer. Consequently
the behavior and radiation property of electron beam can be interpreted upon the
behavior and radiation property of single charged particle. In addition based on the
corrected Lorentz force formula, it is inferred that the angular frequency or velocity
of a charged particle in magnetic dipole is timely independent to its mass but its
total energy status or specifically its internal potential energy status that relates
the charge field's compression status.

\end{document}